\documentclass[sigconf]{acmart}

\usepackage{balance}
\usepackage{url}

\setcopyright{acmcopyright}
\copyrightyear{2020}
\acmYear{2020}

\acmConference[SIGIR '20]{43rd International ACM SIGIR Conference on Research and Development in Information Retrieval}{July 25--30, 2020}{Xi'an, China}
\acmBooktitle{SIGIR 2020}
\acmPrice{15.00}

\begin{document}

\title{Question Answering over Curated and Open Web Sources}

\author{Rishiraj Saha Roy}
\email{rishiraj@mpii.de}
\affiliation{%
  \institution{MPI for Informatics, Germany}
}

\author{Avishek Anand}
\email{anand@l3s.de}
\affiliation{%
  \institution{L3S Research Center, Germany}
}

\begin{abstract}
The last few years have seen an explosion of research on the topic of automated question answering (QA),
spanning the communities of information retrieval, natural language processing, and artificial intelligence.
This tutorial would cover the highlights of this really active period of growth for QA to give the audience a grasp over the families of algorithms that are currently being used. We partition research contributions by the underlying source from where answers are retrieved: curated knowledge graphs, unstructured text, or hybrid corpora.  We choose this dimension of partitioning as it is the most discriminative when it comes to algorithm design. Other key dimensions are covered within each sub-topic: like the complexity of questions addressed, and degrees of explainability and interactivity introduced in the systems. We would conclude the tutorial with the most promising emerging trends in the expanse of QA, that would help new entrants into this field make the best decisions to take the community forward. Much has changed in the community since the last tutorial on QA in SIGIR 2016, and we believe that this timely overview will indeed benefit a large number of conference participants. 
\end{abstract}



\settopmatter{printacmref=false, printccs=false, printfolios=true}


\maketitle

\section{Motivation}
\label{sec:intro}

\subsection{Background}
\label{subsec:background}

Over several decades, the field of question answering (QA) grew steadily from early prototypes like BASEBALL~\cite{green1961baseball}, through IBM Watson~\cite{ferrucci2010building} and all the way to present-day integration in virtually all personal assistants like Siri, Cortana, Alexa, and the Google Assistant. In the last few years though, research on QA has well and truly exploded: this has often resulted in top conferences regularly creating submission tracks and presentation sessions dedicated to this topic. This tutorial will try to highlight key contributions to automated QA systems in the last three to four years coming from the perspectives of information retrieval (IR) and natural language processing (NLP)~\cite{wu2020perq,clark2018simple,christmann2019look,chen2017reading,lu2019answering,guo2018dialog,sun2018open,vakulenko2019message,qiu2020stepwise,dehghani2019learning,rajpurkar2016squad}. 

\subsection{Perspectives}
\label{subsec:perspectives}

In Information Retrieval, QA was traditionally treated as a special use case in search~\cite{voorhees1999trec}, to provide \textit{crisp and direct answers} to certain classes of queries, as an alternative to ranked lists of documents that users would have to sift through. Such queries, with objective answers, are often referred to as \textit{factoid questions}~\cite{clarke2003passage,cucerzan2005factoid} (a term whose definition has evolved over the years). Factoid QA became very popular with the emergence of large curated \textit{knowledge graphs} (KGs) like YAGO~\cite{suchanek2007yago}, DBpedia~\cite{auer2007dbpedia}, Freebase~\cite{bollacker2008freebase} and Wikidata~\cite{vrandevcic2014wikidata}, powerful resources that enable such crisp question answering at scale. \textit{Question answering
over knowledge graphs} or equivalently, knowledge bases (KG-QA or KB-QA) became a field of its own, that is producing an increasing number of research contributions year over year~\cite{vakulenko2019message,wu2020perq,qiu2020stepwise,christmann2019look,shen2019multi,ding2019leveraging,bhutani2019learning}. Effort has also been directed at answering questions over \textit{Web tables}~\cite{pasupat2015compositional,iyyer2017search}, that can be considered canonicalizations of the challenges in QA over structured KGs. 

In contrast, QA (in one of the major senses as we know it today) in Natural Language Processing started with the AI goal of whether machines can comprehend simple passages~\cite{rajpurkar2016squad,yang2018hotpotqa,chen2017reading,clark2018simple} so as to be able to answer questions posed from the contents of these passages. Over time, this machine reading comprehension (MRC) task became coupled with the retrieval pipeline, resulting in the so-called paradigm of \textit{open-domain QA}~\cite{dehghani2019learning,wang2019document,chen2017reading} (a term that is overloaded with other senses as well~\cite{abujabal2018never,elgohary2018dataset}).
Nevertheless, this introduction of the retrieval pipeline led to a revival of \textit{text-QA}, that had increasingly focused on non-factoid QA~\cite{cohen2018wikipassageqa,yang2018query} after the rise of structured KGs. This has also helped bridge the gap between text and KG-QA, with the latter family 
gradually incorporating supplementary textual sources to boost recall~\cite{sun2018open,sun2019pullnet,savenkov2016knowledge}. Considering such heterogeneous sources may often be the right choice owing to the fact that KGs, while capturing an impressive amount of objective world knowledge, are inherently incomplete. 

\vspace{0.2cm}
\noindent \textbf{Terminology.} In this tutorial, we refer to knowledge graphs and Web tables as the \textit{curated Web}, and all unstructured text available online as the \textit{open Web}.

\vspace{0.2cm}
\noindent \textbf{Content.} All material for this tutorial is publicly available at our website \url{https://www.avishekanand.com/talk/sigir20-tute/}.

\section{Objectives}
\label{sec:objectives}

As mentioned in the beginning, the importance of QA has been fuelled to a large extent by the ubiquity of personal assistants: this has also helped bring together these seemingly independent research directions under one umbrella through a unified interface. One of the goals of this tutorial is to give the audience a feel of these commonalities: this can have a significant effect on overcoming the severely fragmented view of the QA community.

\noindent \textbf{What do we not cover?}
QA over relational databases is closely related to the independent field NLIDB (natural language interfaces to databases)~\cite{li2014constructing} and is out of scope of this tutorial. Associated directions like Cloze-style QA~\cite{lewis2019unsupervised} or specialized application domains like biomedical QA~\cite{pampari2018emrqa} will be mentioned cursorily. Approaches for visual and multimodal QA~\cite{guo2019quantifying} are out of scope, and so is community question answering (CQA)
where the primary goal is to match experts with pertinent questions: the answering itself is not by the machine but by humans. 

\section{Relevance to the IR Community}
\label{sec:relevance}

\textbf{Related tutorials.} A tutorial on QA is not really new to SIGIR: the previous one was presented in 2016 by Wen-tau Yih and Hao Ma~\cite{yih2016question} (also at NAACL 2016~\cite{yih2016bquestion}, by the same authors). Text-based QA tutorials appeared way back in NAACL 2001 (Sanda Harabagiu and Dan Moldovan)~\cite{harabagiu2001role}
and EACL 2003 (Jimmy Lin and Boris Katz)~\cite{lin2003question}. Tutorials on IBM Watson~\cite{gliozzo2012natural,fan2015natural} and entity recommendation~\cite{ma2015introduction} have also touched upon QA in the past. Recent workshops on various aspects of QA have been organized at top-tier conferences: MRQA (EMNLP-IJCNLP 2019), RCQA (AAAI 2019), HQA (WWW 2018), and OKBQA (COLING 2016).

\noindent \textbf{Need for a new one.} The unparalleled growth of QA warrants a new tutorial to cover \textit{recent advances} in the field. Primarily a direction dominated by \textit{template-based}~\cite{unger2012template,abujabal2018never} approaches, QA now includes a large number of \textit{neural} methods~\cite{huang2019knowledge,chen2019bidirectional,chen2017reading,clark2018simple}, \textit{graph-based}~\cite{lu2019answering,luo2018knowledge} methods, and even a handful that explore \textit{reinforcement learning}~\cite{qiu2020stepwise,das2018go,pan2019reinforced}. Across sub-fields, more \textit{complex questions} are being handled, complexity being defined in terms of entities and relationships present~\cite{lu2019answering,vakulenko2019message,yang2018hotpotqa}. Systems are moving from their static counterparts to more \textit{interactive} ones: an increasing number of systems are including scope for \textit{user feedback}~\cite{abujabal2018never,zhang2019interactive,kratzwald2019learning} and operate in a \textit{multi-turn, conversational setting}~\cite{shen2019multi,christmann2019look,pan2019reinforced,reddy2019coqa}. \textit{Interpretability} or explainability of presented answers is yet another area of significance~\cite{wu2020perq,abujabal17quint,saeidi2018interpretation,sydorova2019interpretable}, as the role of such explanations is being recognized both for developers and end-users towards system improvement and user satisfaction. The tutorial will emphasize each of these key facets of QA. In addition, a summary  of the available \textit{benchmarks}~\cite{berant2013semantic,rajpurkar2016squad,yang2018hotpotqa,christmann2019look,talmor2018web,choi2018quac,reddy2019coqa,abujabal2019comqa} in each of the QA sub-fields will be provided, that would be very valuable for new entrants to get started with their problem of choice.



\section{Topics}
\label{sec:topics}

\subsection{QA over Knowledge Graphs}
\label{subsec:kg-qa}

The advent of large knowledge graphs like Freebase~\cite{bollacker2008freebase}, YAGO~\cite{suchanek2007yago}, DBpedia~\cite{auer2007dbpedia} and Wikidata~\cite{vrandevcic2014wikidata} gave rise to QA over KGs (KG-QA) that typically provides answers as \textit{single or lists of entities} from the KG. KG-QA has become an important research direction, where the goal is to translate a natural language question into a structured query, typically in the Semantic Web language SPARQL or an equivalent logical form, that directly operates on the entities and predicates of the underlying KG~\cite{wu2020perq,qiu2020stepwise,vakulenko2019message,bhutani2019learning,christmann2019look}.
KG-QA involves challenges of entity disambiguation and, most strikingly, the need to bridge the vocabulary gap between the phrases in a question and the terminology of the KG. Early work on KG-QA built on paraphrase-based mappings and query templates that involve a single entity predicate~\cite{berant2013semantic,unger2012template,yahya2013robust,joshi2014knowledge,yahya2012natural}. This line was further advanced  by~\cite{bast2015more,bao2016constraint,abujabal2017automated,hu2017answering}, including the learning of templates from graph patterns in the KG. However, reliance on templates prevents such approaches from robustly coping with arbitrary syntactic formulations. This has motivated deep learning methods with CNNs and LSTMs, and especially key-value memory networks~\cite{xu2019enhancing,xu2016question,tan2018context,huang2019knowledge,chen2019bidirectional}.

A significant amount in this section of the tutorial will be on answering \textit{complex questions} with multiple entities and predicates. This is one of the key focus areas in KG-QA now~\cite{lu2019answering,bhutani2019learning,qiu2020stepwise,ding2019leveraging,vakulenko2019message,hu2018state,jia18tequila,chakrabarti2020interpretable}, where the overriding principle is often the identification of frequent query substructures. \textit{Web tables} represent a key aspect of the curated Web and contain a substantial volume of structured information~\cite{dong2014knowledge}. QA over such tables contains canonicalized representatives of several challenges faced in large-scale KG-QA, and we will touch upon a few key works in this area~\cite{pasupat2015compositional,iyyer2017search,sun2016table}. 

\subsection{QA over Text}
\label{subsec:text-qa}

\subsubsection{Early efforts} Question answering has originally considered textual document collections as its underlying source. Classical approaches~\cite{ravichandran2002learning,voorhees1999trec} extracted answers from passages and short text units that matched most cue words from the question followed by statistical scoring. This passage-retrieval model makes intensive use of IR techniques for statistical scoring of sentences or passages and aggregation of evidence for answer candidates. TREC ran a QA benchmarking series from 1999 to 2007, and more recently revived it as the LiveQA~\cite{agichtein2015overview} and Complex Answer Retrieval (CAR) tracks~\cite{dietz2017trec}. IBM Watson~\cite{ferrucci2010building} extended this paradigm by combining it with learned models for special question types.

\subsubsection{Machine reading comprehension (MRC)} This is a QA variation where a question needs to be answered as a short \textit{span of words} from a given text paragraph~\cite{rajpurkar2016squad,yang2018hotpotqa}, and is different from the typical fact-centric answer-finding task in IR. Exemplary approaches in MRC that extended the original single-passage setting to a multi-document one can be found in DrQA~\cite{chen2017reading} and DocumentQA~\cite{clark2018simple} (among many, many others). Traditional fact-centric QA over text, and multi-document MRC are recently emerging as a joint topic referred to as open-domain QA~\cite{lin2018denoising,dehghani2019learning,wang2019document}.

\subsubsection{Open-domain QA} In NLP, \emph{open-domain question answering} is now a benchmark task in natural language understanding (NLU) and
can potentially drive the progress of methods in this area~\cite{kwiatkowski2019natural}. The recent reprisal of this task was jump-started by QA benchmarks like SQuAD~\cite{rajpurkar2016squad} and HotpotQA~\cite{yang2018hotpotqa}, that were proposed for MRC. Consequently, a majority of the approaches in NLP focus on MRC-style question answering with varying task complexities~\cite{kwiatkowski2019natural,dasigi2019quoref,talmor2019multiqa,dua2019drop}. This has lead to the common practice of considering the open-domain QA task as a retrieve and re-rank task. In this tutorial, we will introduce the modern foundations of open-domain QA using a similar retrieve-and-rank framework. Note that our focus will not be on architectural engineering but rather on design decisions, task complexity, and the roles and opportunities for IR.

\subsection{QA over Heterogeneous Sources}
\label{subsec:hybrid-qa}

Limitations of QA over KGs has recently led to a revival of considering textual sources, in combination with KGs~\cite{savenkov2016knowledge,xu2016question,sun2018open,sun2019pullnet,sawant2019neural}. Early methods like PARALEX~\cite{fader2013paraphrase} and OQA~\cite{fader2014open} supported noisy KGs in the form of triple spaces compiled via Open IE~\cite{mausam2016open} on Wikipedia articles or Web corpora. TupleInf~\cite{khot2017answering} extended and generalized the Open-IE-based PARALEX approach to complex questions, and is geared for \textit{multiple-choice answer options}. TAQA~\cite{yin2015answering} is another generalization of Open-IE-based QA, by constructing a KG of $n$-tuples from Wikipedia full-text and question-specific search results. SplitQA~\cite{talmor2018web} addressed complex questions by decomposing them into a sequence of simple questions, and relies on crowdsourced training data. Some methods for hybrid QA start with KGs as a source for candidate answers and use text corpora like Wikipedia or ClueWeb as additional evidence~\cite{xu2016question,das2017question,sun2018open,sun2019pullnet,sydorova2019interpretable,xiong2019improving}, or start with answer sentences from text corpora and combine these with KGs for giving crisp entity answers~\cite{sun2015open,savenkov2016knowledge}. 

\subsection{New Horizons in QA}
\label{subsec:emerging}


\subsubsection{Conversational QA} Conversational QA involves a sequence of
questions and answers that appear as a natural dialogue between the system and the user.
The aim of such sequential, multi-turn QA is to understand the context left implicit by users and
effectively answer incomplete and ad hoc follow-up questions. Towards this, various recent benchmarks have been proposed that expect answers that are boolean~\cite{saeidi2018interpretation}, extractive~\cite{choi2018quac} and free-form responses~\cite{reddy2019coqa}, entities~\cite{christmann2019look},
passages~\cite{kaiser2020conversational},
and chit-chat~\cite{zhou2018dataset}. Leaderboards of the QuAC~\cite{choi2018quac} and CoQA~\cite{reddy2019coqa} datasets point to many recent approaches in the text domain.
Recently, the TREC CAsT track~\cite{dalton2019cast}
(\url{http://www.treccast.ai/}) and the Dagstuhl Seminar
on Conversational Search~\cite{anand2020conversational}
tried to address such challenges in \textit{conversational search}.
For KG-QA, notable efforts include~\cite{saha2018complex,guo2018dialog,christmann2019look,shen2019multi}. We will focus on
the modelling complexity of these tasks and a
classification of
the approaches involved. 

\subsubsection{Feedback and interpretability} Static learning systems for QA are gradually paving the way for those that incorporate user \textit{feedback}. These mostly design the setup as a continuous learning setup, where the explicit user feedback mechanism is built on top of an existing QA system. For example, the NEQA~\cite{abujabal2018never}, QApedia~\cite{kratzwald2019learning}, and IMPROVE-QA~\cite{zhang2019interactive} systems primarily operate on the core systems of QUINT~\cite{abujabal2017automated}, DrQA~\cite{chen2017reading}, and gAnswer~\cite{hu2017answering}, respectively. A direction closely coupled with effective feedback is the interpretability of QA models, that is also essential to improving trust and satisfaction~\cite{wu2020perq,abujabal17quint}. This section is for experts and we will discuss potential limitations and open challenges.

\subsubsection{Clarification questions} A key aspect of mixed-initiative systems~\cite{radlinski2017theoretical} is to be able to ask \textit{clarifications}. This ability is essential to QA systems, especially for handling ambiguity and facilitating better question understanding. Most of the work in this domain has
been driven by extracting tasks from open QA forums
~\cite{braslavski2017you,rao2018learning,xu2019asking}.

\section{Format and Support}
\label{sec:format}

A detailed schedule for our proposed \textit{half-day tutorial} (three hours plus breaks), which is aimed to meet a high-quality presentation within the chosen time period, is as follows:

\begin{itemize}
	\vspace{0.2cm}
	\item 9:00 - 10:30 Part I: QA over knowledge graphs (\textbf{1.5 hours})
	\vspace{0.2cm}
		\begin{itemize} 
			\item Background (\textit{15 minutes})
			\item Simple QA (\textit{10 minutes})
			\item Complex QA (\textit{20 minutes})
			\item Heterogeneous QA (\textit{15 minutes})
			\item Conversational QA (\textit{15 minutes})
			\item Summary (\textit{15 minutes})
		\end{itemize}
    \vspace{0.2cm}		
	\item 10:30 - 11:00 Coffee break
	\vspace{0.2cm}
	\item 11:00 - 12:30 Part II: QA over textual sources (\textbf{1.5 hours})
	\vspace{0.2cm}
		\begin{itemize}
			\item Background (\textit{15 minutes})
			\item Machine reading (\textit{20 minutes})
			\item Open-domain QA (\textit{20 minutes})
			\item Feedback and interpretability (\textit{10 minutes})
			\item Conversational QA (\textit{10 minutes})
			\item Summary (\textit{15 minutes})			
		\end{itemize}
\end{itemize}


\section{Presenters}
\label{subsec:bio}

\textbf{Rishiraj Saha Roy} is a Senior Researcher at the Max Planck Institute for Informatics (MPII), Saarbr\"ucken, Germany. He leads the research group on Question Answering, that focuses on robust and interpretable solutions for answering natural language questions over structured and unstructured data. He has more than four years of research experience on question answering. In recent years, he has served on the PCs of conferences like SIGIR, CIKM, AAAI, EMNLP, and NAACL, and published at venues like SIGIR, CIKM, WSDM, WWW, and NAACL.

Prior to joining MPII, he worked for one and a half years as a Computer Scientist at Adobe Research, Bangalore, India. He completed his PhD as a Microsoft Research India Fellow from the Indian Institute of Technology (IIT) Kharagpur. More information about his research can be found at \url{http://people.mpi-inf.mpg.de/~rsaharo/}.

\noindent \textbf{Avishek Anand} is an Assistant Professor at the Leibniz Universit\"at Hannover, Germany. His research interests lie at the intersection of retrieval, mining, and machine learning. He did his PhD at the Max Planck Institute for Informatics, Saarbr\"ucken, Germany, where he worked on temporal information retrieval. Currently, he is developing intelligent and transparent representation learning methods for text and graphs in the context of search, QA and Web tasks. He has published his research in several top-tier conferences, such as SIGIR, WSDM, CIKM, ICDE and TKDE.

He delivered a tutorial at SIGIR 2016, on \emph{Temporal Information Retrieval}, and several talks in various summer schools and invited lectures. Avishek was one of the lead organizers of the recently concluded Dagstuhl seminar on Conversational Search.

\bibliographystyle{ACM-Reference-Format}

\balance

\bibliography{references}

\end{document}